\documentclass[]{svjour2}

\smartqed

\usepackage{amssymb}
\usepackage{graphicx}

\journalname{J Stat Phys}

\begin{document}

\title{Thermodynamics of Two-Component Log-Gases with Alternating Charges}

\titlerunning{Two-component Log-Gases with Alternating Charges}

\author{Ladislav \v{S}amaj}

\institute{Institute of Physics, Slovak Academy of Sciences, 
D\'ubravsk\'a cesta 9, 845 11 Bratislava, Slovakia \\
\email{Ladislav.Samaj@savba.sk}}

\date{Received:  / Accepted: }

\maketitle

\begin{abstract}
We consider a one-dimensional gas of positive and negative unit charges
interacting via a logarithmic potential, which is in thermal equilibrium
at the (dimensionless) inverse temperature $\beta$.
In a previous paper [\v{S}amaj, L.: J. Stat. Phys. {\bf 105}, 173--191 (2001)],
the exact thermodynamics of the unrestricted log-gas of pointlike charges
was obtained using an equivalence with a $(1+1)$-dimensional boundary 
sine-Gordon model. 
The present aim is to extend the exact study of the thermodynamics
to the log-gas on a line with alternating $\pm$ charges. 
The formula for the ordered grand partition function is obtained 
by using the exact results of the Thermodynamic Bethe ansatz.
The complete thermodynamics of the ordered log-gas with pointlike charges
is checked by a small-$\beta$ expansion and at the collapse point $\beta_c=1$.
The inclusion of a small hard core around particles permits us to 
go beyond the collapse point.
The differences between the unconstrained and ordered versions 
of the log-gas are pointed out.


\end{abstract}

\renewcommand{\theequation}{1.\arabic{equation}}
\setcounter{equation}{0}

\section{Introduction} \label{Sect.1}
We study thermal equilibrium properties of a symmetric two-component plasma
(Coulomb gas) which consists of mobile pointlike positive and negative unit 
charges $\pm 1$, confined in a two-dimensional (2D) domain of points 
${\bf r}=(x,y)$.
The charges are immersed in a homogeneous medium of dielectric constant 1,
the system as a whole is electroneutral.
In Gauss units, the Coulomb potential $\phi({\bf r})$, induced by 
a unit charge at the origin ${\bf 0}$, is given by the 2D Poisson equation
\begin{equation} \label{Poisson}
\Delta \phi({\bf r}) = - 2\pi \delta({\bf r}) . 
\end{equation}
Such definition of the 2D Coulomb potential maintains many generic
properties of ``realistic'' three-dimensional Coulomb systems with 
$1/r$ potential, e.g. the screening sum rules \cite{Martin88}.
In an infinite space, the solution of Eq. (\ref{Poisson}) reads
\begin{equation} \label{Coulombpotential}
\phi({\bf r}) = - \ln \left( \frac{r}{r_0} \right) ,
\end{equation}
where $r=\vert {\bf r}\vert$ and $r_0$ is a length scale which fixes 
the zero point of the potential.
The interaction energy of two charges $q$ and $q'$ at the respective
spatial positions ${\bf r}$ and ${\bf r}'$ is equal to 
$v_{qq'}({\bf r},{\bf r}') = q q' \phi(\vert {\bf r}-{\bf r}'\vert)$.
The corresponding Boltzmann factor at the (dimensionless) inverse temperature 
$\beta=1/(k_{\rm B}T)$ reads $\exp[-\beta v_{qq'}({\bf r},{\bf r}')] 
= \vert {\bf r}-{\bf r}'\vert^{\beta q q'}$.
For two oppositely charged pointlike species at distance $r$, 
the Boltzmann factor $r^{-\beta}$ is integrable at small distances 
$r\to 0$ for small enough $\beta$, $\beta<\beta_c$.
Thus, the thermodynamics of pointlike charges on a continuum space
is well defined only in the high-temperature region $\beta<\beta_c$. 
For low temperatures $\beta\ge \beta_c$, the short-distance (ultraviolet) 
collapse of oppositely charged point particles makes the thermodynamics 
unstable; the thermodynamic stability is restored by considering 
a short-distance regularization of the Coulomb potential, i.e. via hard 
cores attached to particles which prevent them from touching one another.
The value of the collapse (inverse) temperature $\beta_c$ depends on the
dimensionality of the compact domain $\Lambda$ to which the charge system is 
constrained.

In the case of a 2D domain $\Lambda$, the 2D integral 
$\int_{\Lambda} {\rm d}^2r\, \vert r\vert^{-\beta}$ is finite at small distances 
provided that $\beta<2$, i.e. $\beta_c=2$; there is no problem at large 
$\vert r\vert$ because the interaction is screened by the conducting system.
A typical domain of this kind is an infinite space, $\Lambda\to \mathbb{R}^2$,
which defines the bulk 2D Coulomb gas.
The complete thermodynamics of the system of pointlike $\pm$ charges in 
the stability region $\beta<2$ was derived in \cite{Samaj00}; 
for a short review of exact results for bulk and surface thermodynamics, 
together with asymptotic large-distance behavior of charge and density 
correlation functions, see \cite{Samaj03}.
The derivation was based on an equivalence between the 2D Coulomb gas and 
the (1+1)-dimensional sine-Gordon theory, transferring from that integrable 
theory some exact results obtained within 
the Thermodynamic Bethe ansatz (TBA). 
The extension of the exact thermodynamics of the bulk Coulomb gas beyond 
the collapse point $\beta_c=2$, with a hard-core regularization of 
the Coulomb interaction and in the region of low particle densities, 
was proposed in \cite{Kalinay02}.
An electroneutrality sum rule and the leading short-distance behavior of
pair correlation functions was used to go up to $\beta=3$.
Applying a systematic short-distance expansion of correlation functions 
\cite{Tellez05}, T\'ellez showed \cite{Tellez07} that one can proceed, 
in principle, up to the Kosterlitz-Thouless transition of an infinite order
from a high-temperature conductor phase (a non-zero fraction of the positive 
and negative charges is dissociated) to a low-temperature insulator phase
(the positive and negative charges form neutral pairs) which takes place at 
$\beta_{\rm KT}=4$ for low densities.

In the case of a one-dimensional (1D) domain $\Lambda$, say the infinite line 
$x\in (-\infty,\infty)$, the 1D integral $\int_{\Lambda} {\rm d}x\, x^{-\beta}$ 
is finite at small distances provided that $\beta<1$, i.e. $\beta_c=1$.
Such systems, to which we refer as log-gases, have evoked much of interest 
because of their relationship to various models of condensed matter. 
There exist two basic versions of two-component log-gases: 
without and with space restriction on the ordering of $\pm$ charges.
\begin{itemize}
\item
{\bf Unconstrained log-gas:}
The ``standard'' 1D log-gas without any space restriction on the $\pm$ 
charges is related to dissipative quantum mechanics 
\cite{Schmid83,Caldeira83,Guinea85,Callan92} and to the problem of 
non-equilibrium quantum transport through a point contact in a 1D Luttinger 
liquid \cite{Kane92,Fendley95b}.
The lattice version of the model, which represents a kind of short-distance
regularization of the Coulomb potential, was exactly solved 
(the grand partition function and the particle correlation functions) at 
$\beta=1$ \cite{Forrester86b} and $\beta=2,4$ \cite{Forrester88,Forrester89}.
The conductor-insulator phase diagram was conjectured in \cite{Forrester90};
various approaches indicate that the Kosterlitz-Thouless phase transition
should occur at $\beta_{\rm KT}=2$, independently of the particle density.
Thermodynamic properties of the continuous version of the model with
pointlike charges was solved in the whole stability region $\beta<1$
in \cite{Samaj01}, by exploring the TBA results for an equivalent
$(1+1)$-dimensional boundary sine-Gordon theory 
\cite{Fendley95a,Fendley95b,Bazhanov97,Fateev97}.
\item
{\bf Log-gas with charge ordering:}
The 1D log-gas with $\pm$ charges required to alternate in space
is equivalent to the Kondo problem with spin-$\frac{1}{2}$ impurity
\cite{Anderson69,Anderson70,Schotte71}.
The lattice version of the system is exactly solvable at the ``collapse''
isotherm $\beta=1$ \cite{Forrester86b}.
The charge ordered system is expected to exhibit a dielectric phase
at an arbitrary temperature \cite{Forrester86a}.

\end{itemize}
The asymptotic large-distance behavior of the particle correlation functions 
was studied in detail for both versions of the 1D two-component log-gas
in \cite{Alastuey95}. 

The aim of the present work is to extend the exact study of the thermodynamics
to the two-component log-gas on a line with alternating $\pm$ charges. 
The basic formula for the ordered grand partition function is obtained 
by using the TBA results \cite{Fendley95c,Fendley96}, associated with
a fusion relation between the grand partition functions of the unconstrained
and ordered log-gases. 
The complete thermodynamics of the ordered log-gas with pointlike charges
is derived.
The results are checked by a small-$\beta$ expansion and at the
collapse point $\beta_c=1$.
The inclusion of small hard cores around particles permits us to extend 
the thermal analysis beyond the collapse point $\beta_c=1$.
The important differences between the unconstrained and ordered versions
of the log-gas are pointed out.

The paper is organized as follows. 

We start with a brief recapitulation and an extension of the results
for the unconstrained 1D log-gas \cite{Samaj01} in Sect. 2.
Sect. 2.1 deals with the case of pointlike charges. 
The inclusion of a small hard core around particles, which permits us
to study the thermodynamics beyond the collapse point $\beta_c=1$, is 
the subject of Sect. 2.2.
Here, we apply two methods: the one based on a perfect screening sum rule 
valid for conducting systems and the other, more general, based on 
the explicit definition of the grand partition function.

The thermodynamics of the ordered 1D log-gas is derived in Sect. 3.
As in the previous unconstrained case, the system of pointlike charges is 
treated in Sect. 3.1 and the inclusion of the hard core to particles, 
via the definition of the grand partition function, is worked out in Sect. 3.2.
The differences between the unconstrained and ordered versions of the log-gas 
are pointed out.   

Sect. 4 is Conclusion.

Auxiliary calculations are shifted aside to Appendices.
A fusion relation between the grand partition functions of the unconstrained
and ordered log-gases is used to rederive in an alternative way a relationship 
between the corresponding bulk pressures in Appendix A.
The small-$\beta$ expansion of the obtained density-fugacity relationship 
for the ordered log-gas is checked by microscopic calculations in Appendix B.

\renewcommand{\theequation}{2.\arabic{equation}}
\setcounter{equation}{0}

\section{Extension of the results for the unconstrained log-gas} \label{Sect.2}
First we recapitulate and extend the exact results for the thermodynamics of
the unconstrained log-gas obtained in \cite{Samaj01}.
We start with pointlike charges.
The inclusion of hard cores to particles, which will allow us to pass
through the collapse point $\beta_c=1$, is described in the subsequent part.

\subsection{Pointlike particles}
We go to an infinite line through the circle of radius $R$ and
circumference $L=2\pi R$, taking at the end the limit $R\to\infty$.
The position of a particle on the circle is specified by
the angle $\varphi\in[0,2\pi)$.
Since the distance between particles at angle positions $\varphi$
and $\varphi'$ is $2R\vert \sin[(\varphi-\varphi')/2]\vert$,
the 2D Coulomb potential takes the form
\begin{equation}
\phi(\varphi,\varphi') = - \ln \left[ \frac{2R}{r_0}
\left\vert \sin\left( \frac{\varphi-\varphi'}{2} \right) \right\vert \right] .
\end{equation} 
Defining the line coordinate as $x=R\varphi$, the 1D realization of
the potential (\ref{Coulombpotential}), 
$\phi(x,x') = -\ln(\vert x-x'\vert/r_0)$,
results from this expression as the $R\to\infty$ limit.
The Boltzmann factor of two charges $q$ and $q'$ at the respective angle 
positions $\varphi$ and $\varphi'$ reads as
\begin{equation}
\exp\left[ -\beta q q' \phi(\varphi,\varphi') \right]
= \left\vert \frac{2R}{r_0} \sin\left( \frac{\varphi-\varphi'}{2} \right)
\right\vert^{\beta qq'} .
\end{equation}

We shall work in the grand canonical ensemble.
Due to the charge $\pm$ symmetry, chemical potentials of the species 
can be taken equivalent, $\mu_+=\mu_-=\mu$.
Only neutral configurations with the particle numbers $N_+=N_-=N$ are 
considered. 
The position angles of $(+)$ charges will be denoted by $\varphi_i$ 
and those of $(-)$ charges by $\varphi'_i$ $(i=1,\ldots,N)$. 
The grand partition function is defined as
\begin{eqnarray}
\Xi_L(\mu) & = & \sum_{N=0}^{\infty} \frac{\exp(2\beta\mu N)}{(N!)^2}
\int_0^{2\pi} \frac{{\rm d}\varphi_1 R}{\lambda}
\int_0^{2\pi} \frac{{\rm d}\varphi'_1 R}{\lambda} \cdots
\int_0^{2\pi} \frac{{\rm d}\varphi_N R}{\lambda}
\int_0^{2\pi} \frac{{\rm d}\varphi'_N R}{\lambda} \nonumber \\ 
& & \times \left\vert \frac{\prod_{(i<j)=1}^N 
\left[\frac{2R}{r_0} \sin\left( \frac{\varphi_i-\varphi_j}{2} \right)\right]
\left[\frac{2R}{r_0} \sin\left( \frac{\varphi'_i-\varphi'_j}{2} \right)\right]}{
\prod_{i,j=1}^N \left[\frac{2R}{r_0} \sin\left( \frac{\varphi_i-\varphi'_j}{2} 
\right)\right]} \right\vert^{\beta} , \label{Xiorig}
\end{eqnarray}
where $\lambda$ is the de Broglie wavelength and the dependence on 
the inverse temperature $\beta$ is omitted.
We introduce the rescaled fugacity $z_+=z_-=z$,
\begin{equation} \label{fugacity}
z = \exp(\beta\mu) \frac{r_0^{\beta/2}}{\lambda} ,
\end{equation}
which has the dimension $[{\rm length}]^{\frac{\beta}{2}-1}$.
Factorizing out quantities of nonzero dimensions in (\ref{Xiorig}),
we end up with the representation
\begin{equation} \label{Xitrans}
\Xi_L(z) = 1 + \sum_{N=1}^{\infty} \left[ (2\pi)^{\beta/2} z L^{1-\frac{\beta}{2}}  
\right]^{2N} I_{2N} ,
\end{equation}
where the dimensionless configuration integrals
\begin{equation} \label{defI}
I_{2N} = \frac{1}{(N!)^2} \int_0^{2\pi} \frac{{\rm d}\varphi_1}{2\pi} 
\int_0^{2\pi} \frac{{\rm d}\varphi'_1}{2\pi} \cdots
\int_0^{2\pi} \frac{{\rm d}\varphi_N}{2\pi}  
\int_0^{2\pi} \frac{{\rm d}\varphi'_N}{2\pi} 
B_{2N}(\{ \varphi_i\},\{ \varphi'_i\}) 
\end{equation}
involve the Boltzmann weights
\begin{equation} \label{Boltzmannweight}
B_{2N}(\{ \varphi_i\},\{ \varphi'_i\}) = \left\vert \frac{
\prod_{(i<j)=1}^N \left[2\sin\left( \frac{\varphi_i-\varphi_j}{2} \right)\right]
\left[2\sin\left( \frac{\varphi'_i-\varphi'_j}{2} \right)\right]}{
\prod_{i,j=1}^N \left[2\sin\left( \frac{\varphi_i-\varphi'_j}{2} 
\right)\right]} \right\vert^{\beta} .
\end{equation}

In the large-$L$ limit, we pass from a finite circle to an infinite
1D line, $x\in (-\infty,\infty)$.   
According to elementary thermodynamics, the quantity $\ln \Xi_L(z)$ 
is expected to be extensive. 
In particular, the bulk pressure of the charge system $P(z)$, defined by
\begin{equation} \label{defpressure}
\beta P(z) = \lim_{L\to\infty} \frac{1}{L} \ln \Xi_L(z) ,
\end{equation}
has a well-defined finite value.
The only expansion parameter in the series representation (\ref{Xitrans}) 
is the dimensionless combination $z L^{1-\frac{\beta}{2}}$, so that
$\beta P(z) \propto z^{\frac{2}{2-\beta}}$.
The proportionality $\beta$-dependent factor was found in \cite{Samaj01} 
by using the TBA results obtained for an equivalent boundary sine-Gordon model
\cite{Fendley95a,Fendley95b,Bazhanov97,Fateev97}.
We take the TBA results from \cite{Fendley95a}; the notation in that
work is related to ours as follows: $g=\beta/2$, $t=1/g=2/\beta$,
$q=\exp({\rm i}\pi g) = \exp({\rm i}\pi\beta/2)$, $T=1/L$ and a scale
$T_{\rm B}=1/L_{\rm B}$.
The expansion parameter in (\ref{Xitrans}) was expressed in terms
of the length scale $L_{\rm B}$ as
\begin{equation} \label{exp}
z (2\pi)^{\beta/2} L^{1-\frac{\beta}{2}} = \Gamma\left(\frac{\beta}{2}\right)
\left[ \frac{L}{L_{\rm B}} \frac{\Gamma\left(\frac{1}{2-\beta}\right)}{
2\sqrt{\pi}\Gamma\left(\frac{\beta}{2(2-\beta)}\right)} 
\right]^{1-\frac{\beta}{2}} ,
\end{equation}
where $\Gamma$ is the Gamma function \cite{Gradshteyn}.
For large $L$, the grand partition function is given by
\begin{equation}
\Xi_L \mathop{\sim}_{L\to\infty} \sqrt{\frac{\beta}{2}} \exp\left[ 
\frac{L}{L_{\rm B}} \frac{1}{2\cos\left(\frac{\pi\beta}{2(2-\beta)}\right)} 
\right] .
\end{equation}
Eliminating $L_{\rm B}$ from the couple of equations and using formulas 
for the Gamma functions \cite{Gradshteyn}
\begin{equation} \label{Gamma}
\Gamma(1-y) \Gamma(y) = \frac{\pi}{\sin(\pi y)} , \qquad
\Gamma\left( \frac{1}{2}-y\right) \Gamma\left(\frac{1}{2}+y\right) 
= \frac{\pi}{\cos(\pi y)}
\end{equation}
with real $y$, we finally arrive at the bulk pressure
\begin{equation} \label{betaP}
\beta P(z) = \frac{1}{2\pi^{3/2}} \Gamma\left( \frac{1-\beta}{2-\beta} \right)
\Gamma\left( \frac{\beta}{2(2-\beta)} \right) 
\left[ \frac{2\pi z}{\Gamma(\beta/2)} \right]^{\frac{2}{2-\beta}} .
\end{equation}

The one-body densities are defined as thermal averages
\begin{equation}
n_q(x) \equiv n_q = \left\langle \sum_j \delta_{q,q_j} \delta(x-x_j) 
\right\rangle ,  
\end{equation}
where $\delta_{q,q'}$ is the Kronecker symbol, $\delta(x-x')$ the Dirac
delta function and the index $j$ runs over all particles.
For the considered charge $\pm$ symmetry, the species densities are equal
to one another: $n_+=n_-=n/2$, where the total number density of particles 
$n$ is given by
\begin{equation} \label{nP}
n(z) = z \frac{\partial}{\partial z} \beta P(z) .
\end{equation}
Thus the density-fugacity relationship reads as
\begin{equation} \label{nz1}
\frac{n}{z^{\frac{2}{2-\beta}}} = \frac{1}{\pi^{3/2}(2-\beta)} 
\Gamma\left( \frac{1-\beta}{2-\beta} \right)
\Gamma\left( \frac{\beta}{2(2-\beta)} \right) 
\left[ \frac{2\pi}{\Gamma(\beta/2)} \right]^{\frac{2}{2-\beta}} .
\end{equation}

The small-$\beta$ expansion of the rhs of this formula 
\begin{equation} \label{smallb}
\frac{n}{z^{\frac{2}{2-\beta}}} = 2 \beta^{\frac{\beta}{2-\beta}} \exp\left( 
\left[ C + \ln (2\pi) \right] \frac{\beta}{2} + O(\beta^2) \right) ,
\end{equation}
where $C$ is the Euler constant, was checked in Appendix of 
\cite{Samaj01} by using a renormalized Mayer expansion.
Note that this expansion is non-analytic due to the appearance of
the term $\beta^{\frac{\beta}{2-\beta}}$.
On the other hand, the series in the exponential is analytic in $\beta$. 

For a fixed $z$ and in the limit $\beta\to 1^-$, the term 
$\Gamma((1-\beta)/(2-\beta))\sim 1/(1-\beta)$ implies that the particle 
density $n$ exhibits the expected collapse singularity
\begin{equation} \label{sing}
n \mathop{\sim}_{\beta\to 1^-} \frac{4 z^2}{1-\beta} . 
\end{equation}
This singular behavior can be deduced indirectly by using a perfect
screening sum rule for the one-body densities of the conducting 
system \cite{Martin88},
\begin{equation} \label{screening}
n_q = \int {\rm d}x\, \left[ U_{q,-q}(x) - U_{q,q}(x) \right] ,
\end{equation}
where the Ursell functions are defined by
\begin{equation} \label{short}
U_{q,q'}(x,x') \equiv U_{q,q'}(\vert x-x'\vert) 
= \left\langle \sum_{j\ne k} \delta_{q,q_j} \delta(x-x_j) 
\delta_{q',q_k} \delta(x'-x_k) \right\rangle - n_q n_{q'} .
\end{equation} 
The short-distance behavior of the Ursell function for a positive-negative
pair of charges is given by the Boltzmann factor of the pair interaction
\cite{Jancovici77,Hansen85},
\begin{equation} \label{shortdist}
U_{q,-q}(x) \mathop{\sim}_{x\to 0} z^2 \vert x\vert^{-\beta} .
\end{equation}
For $\beta\to 1^-$, the integral in (\ref{screening}) is dominated by
this short-distance behavior and we have
\begin{equation}
\frac{n}{2} \sim \int_{-\ell}^{\ell} {\rm d}x\, 
\frac{z^2}{\vert x\vert^{\beta}} = \frac{2 z^2}{1-\beta} \ell^{1-\beta}
= \frac{2 z^2}{1-\beta} + O(1) ,
\end{equation}
where $\ell$ is a screening length of the Coulomb potential.  
This derivation of the singular behavior (\ref{sing}) points out
the two-body nature of the collapse phenomenon. 

\begin{figure}[h] \label{fig1} 
\begin{center}
\includegraphics[clip,width=0.85\textwidth]{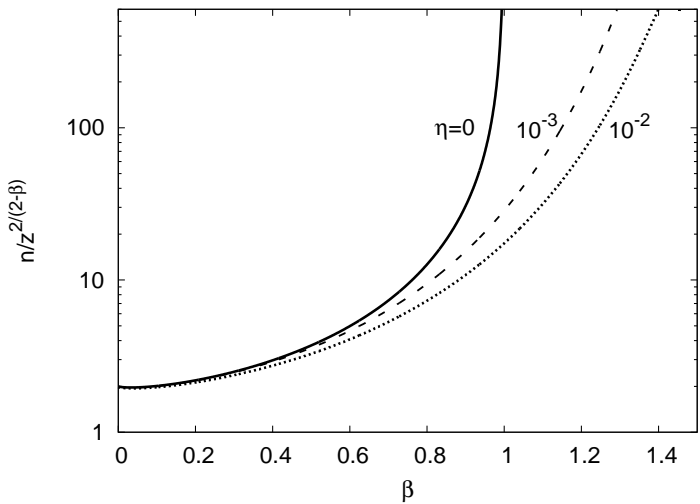} 
\end{center}
\caption{The density-fugacity relationship versus the inverse temperature 
$\beta$ for the unconstrained log-gas.
The packing fraction $\eta=n\sigma$, where $\sigma$ is the hard-core diameter.
The solid curve describes pointlike charges $(\eta=0)$. 
The dashed and dotted curves correspond to $\eta=10^{-3}$ and
$10^{-2}$, respectively.}
\end{figure}

The plot of the density-fugacity relationship for pointlike particles
(\ref{nz1}), well defined up to $\beta_c=1$, is represented in Fig. 1 
by the solid curve. 
Although the function looks like monotonously increasing, this is not true:
as is seen in Fig. 2, the function decreases and reaches the minimum
for small $\beta$. 

\begin{figure}[h] \label{fig2} 
\begin{center}
\includegraphics[clip,width=0.85\textwidth]{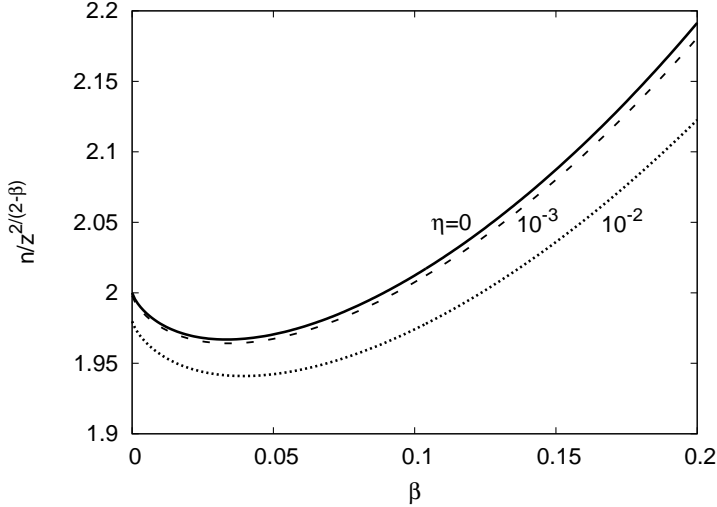} 
\end{center}
\caption{Fig. 1 on a smaller scale to document the non-monotonous
behavior of the density-fugacity plot for small $\beta$.}
\end{figure}

The virial equation of state, which relates the pressure $P$ and the particle 
density $n$, takes a simple form 
\begin{equation}
\beta P = \left( 1 - \frac{\beta}{2} \right) n 
\end{equation}
which is characteristic for log-gas systems. 

Having the density-fugacity relationship, the full thermodynamics is
obtained by passing from the grand-canonical to the canonical ensemble
via the Legendre transformation
\begin{equation}
F_L(\beta,N) = - \frac{1}{\beta} \ln\Xi_L + \mu N ,
\end{equation}
where $F_L$ is the Helmholtz free energy and $N=n L$.
The (dimensionless) specific free energy, defined as $f\equiv \beta F_L/N$, 
thus reads
\begin{eqnarray}
f(\beta,n) & = & \ln(\lambda n) - \frac{\beta}{2} \ln(2r_0 n)
+ \frac{1}{2} \left( 1 - \frac{3\beta}{2}\right) \ln\pi 
\nonumber \\ & & - \left( 1 - \frac{\beta}{2} \right) \left[ 
1 - \ln\left( 1 - \frac{\beta}{2} \right) \right] 
+ \ln\Gamma\left( \frac{\beta}{2}\right) \nonumber \\ & & 
- \left( 1 - \frac{\beta}{2} \right) \left[  
\ln\Gamma\left( \frac{1-\beta}{2-\beta} \right)
+ \ln\Gamma\left( \frac{\beta}{2(2-\beta)} \right) \right] .
\end{eqnarray}
The excess (i.e. over ideal) internal energy per particle and specific heat 
at constant volume are given by
\begin{equation}
u^{\rm ex} = \frac{\partial}{\partial \beta} f(\beta,n) , \qquad
\frac{c_v^{\rm ex}}{k_{\rm B}} = 
-\beta^2 \frac{\partial^2}{\partial \beta^2} f(\beta,n) ,
\end{equation}
respectively.
For the specific heat, we obtain explicitly
\begin{eqnarray}
\frac{c_v^{\rm ex}}{k_{\rm B}} & = & \frac{\beta^2}{2(2-\beta)^3}
\left[ \psi'\left(\frac{1-\beta}{2-\beta}\right) +
\psi'\left(\frac{\beta}{2(2-\beta)}\right) \right] \nonumber \\
& & - \frac{\beta^2}{2(2-\beta)} 
- \frac{\beta^2}{4} \psi'\left(\frac{\beta}{2}\right)  ,
\end{eqnarray}
where the psi function and its derivative are given by
\begin{equation}
\psi(x) = \frac{{\rm d}}{{\rm d}x} \ln\Gamma(x) , \qquad
\psi'(x) = \sum_{j=0}^{\infty} \frac{1}{(x+j)^2} .
\end{equation}
The specific heat exhibits the Laurent series expansion around 
the collapse point $\beta\to 1^-$, 
\begin{equation} \label{specsing}
\frac{c_v^{\rm ex}}{k_{\rm B}} = \frac{1}{2(1-\beta)^2}
- \frac{3}{2(1-\beta)} + O(1) .
\end{equation}

\subsection{Particles with hard cores}
To go beyond the collapse point $\beta_c=1$, we have to consider
a short-distance regularization of the interaction Coulomb potential,
especially for oppositely charged particles.
We introduce a hard core of diameter $\sigma$ around each particle which 
does not allow the charges to touch one another:
\begin{equation} \label{hardcore}
v_{q,q'}(x,x') = \left\{
\begin{array}{lll}
- q q' \ln(\vert x-x'\vert/r_0) & & \mbox{if $\vert x-x'\vert>\sigma$,} \cr 
& \cr \infty & & \mbox{if $\vert x-x'\vert\le \sigma$.}
\end{array} \right.
\end{equation}
The total particle density now depends also on $\sigma$, $n(z,\sigma)$.
The hard-core diameter $\sigma$, or more precisely the dimensionless 
combination 
\begin{equation} \label{defxi}
\xi = \sigma z^{\frac{2}{2-\beta}} ,
\end{equation} 
will be assumed to be small, $\xi\to 0$, and we shall look for 
the leading hard-core correction to the pointlike system.

We first derive the leading hard-core correction in analogy with 
the derivation for the 2D Coulomb gas presented in \cite{Kalinay02}.
The unconstrained log-gas system is in its conducting phase up to
$\beta_{\rm KT}=2$, so that the screening sum rule (\ref{screening}) holds 
and the Ursell functions are well defined also beyond the collapse point.
The difference $U_{q,-q}-U_{q,q}$ vanishes inside the hard core for the
potential (\ref{hardcore}), and the total particle number density
is given by
\begin{equation} \label{newscreening}
n(z,\sigma) = 2 \int_{\sigma}^{\infty} {\rm d}x\,
\left[ U_{q,-q}(x;z,\sigma) - U_{q,q}(x;z,\sigma) \right] .
\end{equation}
We can write
\begin{equation}
U_{q,q'}(x;z,\sigma) = U_{q,q'}(x;z,0) + \Delta_{q,q'}(x;z,\sigma) ,
\qquad x>\sigma , 
\end{equation}
which defines $\Delta_{q,q'}(x;z,\sigma)$, vanishing when $\sigma\to 0$,
as the change of the Ursell function due to the introduction of the hard core
$\sigma$ to pointlike particles. 
Subtraction of Eq. (\ref{newscreening}) with $\sigma>0$ and the same equation
with $\sigma=0$ leads to
\begin{eqnarray}
n(z,\sigma) - n(z,0) & = & - 2 \int_0^{\sigma} {\rm d}x\,
\left[ U_{q,-q}(x;z,0) - U_{q,q}(x;z,0) \right] \nonumber \\
& & + 2 \int_{\sigma}^{\infty} {\rm d}x\,
\left[ \Delta_{q,-q}(x;z,\sigma) - \Delta_{q,q}(x;z,\sigma) \right] .
\label{subtraction}
\end{eqnarray}
We make a heuristic assumption: in the small-$\sigma$ limit, we can
neglect the quantities $\Delta_{q,\pm q}$ in (\ref{subtraction}) as
they contribute to subleading hard-core corrections.
Such assumption has already been made in \cite{Cornu87,Cornu89,Kalinay02}
for the 2D Coulomb gas.
Consequently,
\begin{equation} \label{screen}
n(z,\sigma) = n(z,0) - 2 \int_0^{\sigma} {\rm d}x\,
\left[ U_{q,-q}(x;z,0) - U_{q,q}(x;z,0) \right] . 
\end{equation}
As before, the integral is dominated by the short-distance behavior
of the Ursell function (\ref{shortdist}) and we arrive at the basic result
\begin{equation} \label{basic}
n(z,\sigma) = n(z,0) - 4 z^2 \frac{\sigma^{1-\beta}}{1-\beta} ,
\qquad \xi \to 0 .
\end{equation}
Strictly speaking, this result was derived in the stability region 
$0\le \beta <1$ where $n(z,0)$ is well defined by (\ref{nz1}).
It is however reasonable to assume that the formula (\ref{basic}) can be
analytically continued beyond $\beta_c=1$ because both 
the sum rule (\ref{newscreening}) and the leading short-distance behavior 
(\ref{shortdist}), which were crucial in its derivation, 
remain valid up to $\beta_{\rm KT}=2$.

To provide a convenient representation of (\ref{basic}), we express 
the density-fugacity relation for pointlike particles (\ref{nz1}) in the form
\begin{equation} \label{pointdensity}
n(z,0) = \frac{4\Phi(\beta)}{1-\beta} z^{\frac{2}{2-\beta}} ,
\end{equation}
where the function
\begin{equation} \label{Phi}
\Phi(\beta) = \frac{1}{4\pi^{3/2}} 
\Gamma\left( \frac{3-2\beta}{2-\beta} \right)
\Gamma\left( \frac{\beta}{2(2-\beta)} \right)
\left[ \frac{2\pi}{\Gamma(\beta/2)} \right]^{\frac{2}{2-\beta}} 
\end{equation}
results from the equality $\Gamma(x+1)=x\Gamma(x)$ for $x=(1-\beta)/(2-\beta)$.
$\Phi(\beta)$ was chosen such that it equals to 1 at the collapse point, namely
\begin{equation}
\Phi(\beta) = 1 + \left[ C +\ln(4\pi) \right] (\beta-1) + O((\beta-1)^2) .
\end{equation}
Using the representation (\ref{pointdensity}) in (\ref{basic}), we get
the density-fugacity relationship which involves the leading hard-core 
correction:
\begin{equation} \label{hcresult}
\frac{n(z,\sigma)}{z^{\frac{2}{2-\beta}}} = \frac{4}{1-\beta}
\left[ \Phi(\beta) - \xi^{1-\beta} \right] , \qquad \xi\to 0 .
\end{equation}
The leading correction $\propto \xi^{1-\beta}$ is marginal in the stability
region $0\le \beta <1$.
It makes the density finite at the collapse point $\beta_c=1$,
\begin{equation}
\frac{n(z,\sigma)}{z^2} = - 4 \left[ \ln\xi + C + \ln(4\pi) \right] .
\end{equation}
Note that the density is positive for small $\xi$ and diverges for
$\xi\to 0$, as it should be.
The leading correction becomes relevant in the region $\beta\ge 1$,
up to $\beta=3/2$ at which $\Phi(\beta)$ diverges.
To overcome this divergence, we have to account for higher-order terms of 
the short-distance expansion of the Ursell functions in the screening sum 
rule (\ref{screen}), in close analogy with \cite{Kalinay02,Tellez05,Tellez07}. 
Such analysis goes beyond the scope of this work.

It is convenient to pass from the parameter $\xi$ to the packing fraction
\begin{equation}
\eta = n \sigma .
\end{equation}
Since each particle occupies the length $\sigma$, it must hold that
$L/N\ge \sigma$, i.e. $\eta\le 1$.
Denoting $f\equiv n(z,\sigma)/z^{\frac{2}{2-\beta}}$, we have $\eta = f\xi$.
Multiplying both sides of Eq. (\ref{hcresult}) by 
$(1-\beta) f^{1-\beta}$, we get
\begin{equation}
(1-\beta) f^{2-\beta} - 4 \left[ \Phi(\beta) f^{1-\beta} - \eta^{1-\beta}
\right] = 0 .
\end{equation}
At $\beta=0$, from two solutions of this equation we take the one
$f=1+\sqrt{1-4\eta}$ and follow this branch when increasing $\beta$.
The density-fugacity relation for the packing fractions $10^{-3}$ and
$10^{-2}$ are plotted in Figs. 1 and 2 by the dashed and dotted curves, 
respectively.

For particles with hard cores, in analogy with (\ref{nP}) it holds  
\begin{equation} 
n(z,\sigma) = z \frac{\partial}{\partial z} \beta P(z,\sigma) .
\end{equation}
With regard to (\ref{basic}), we obtain the equation of state
\begin{equation} \label{resultpressure}
\beta P(z,\sigma) = \beta P(z,0) - 2 z^2 \frac{\sigma^{1-\beta}}{1-\beta} ,
\qquad \xi\to 0 ,
\end{equation}
where $\beta P(z,0)$ is given by (\ref{betaP}).
This relation can be deduced directly from the definition of the grand
partition function; the importance of such derivation is evident for
non-conducting systems which do not exhibit screening.
As the collapse phenomenon does not involve $(++)$ and $(--)$ pairs of charges,
we simplify the model by considering hard cores only between oppositely 
charged particles.
In particular, $v_{qq}(x,x')=-\ln(\vert x-x'\vert/r_0)$ for an arbitrary
distance between particles and
\begin{equation} 
v_{q,-q}(x,x') = \left\{
\begin{array}{lll}
\ln(\vert x-x'\vert/r_0) & & \mbox{if $\vert x-x'\vert>\sigma$,} \cr 
& \cr \infty & & \mbox{if $\vert x-x'\vert\le \sigma$.}
\end{array} \right.
\end{equation}
With the definition (\ref{fugacity}) of the fugacity $z$, the grand
partition function of this neutral 1D system of size $L$ (which is large 
enough to neglect boundary effects, but not infinite) is given by
\begin{equation} \label{xinew}
\Xi_L(z,\sigma) = 1 + \sum_{N=1}^{\infty} z^{2N} I_{2N}(\sigma) ,
\end{equation}
where the configuration integrals
\begin{eqnarray} 
I_{2N}(\sigma) & = & \frac{1}{(N!)^2} \int_0^L {\rm d}x_1 \int_0^L {\rm d}x'_1
\cdots \int_0^L {\rm d}x_N \int_0^L {\rm d}x'_N \nonumber \\ & & \times
B_{2N}(\{ x_i\},\{ x'_i\}) \prod_{i,j=1}^N \theta(\vert x_i-x'_j\vert-\sigma)
\label{confint}
\end{eqnarray}
involve the Boltzmann weights
\begin{equation} \label{Boltzmann}
B_{2N}(\{ x_i\},\{ x'_i\}) = \left\vert \frac{\prod_{(i<j)=1}^N 
(x_i-x_j)(x'_i-x'_j)}{\prod_{i,j=1}^N (x_i-x'_j)} \right\vert^{\beta} .
\end{equation}
Here, $\theta(x)$ is the Heaviside theta function:
\begin{equation}
\theta(x) = \left\{ 
\begin{array}{ll}
1 & \mbox{if $x>0$,} \cr
0 & \mbox{if $x\le 0$,}
\end{array} \right.
\end{equation}
which fulfills the obvious relation $\theta(x)+\theta(-x)=1$.
Our trick consists in substituting each theta function in (\ref{confint}) by
\begin{equation} \label{trick}
\theta(\vert x_i-x'_j\vert-\sigma) = 1 - \theta(\sigma-\vert x_i-x'_j\vert) .
\end{equation} 
For the lowest-order integral $I_2(\sigma)$, we get
\begin{eqnarray}
I_2(\sigma) & = & I_2(0) - \frac{1}{(1!)^2} \int_0^L {\rm d}x_1 \int_0^L 
{\rm d}x'_1 \frac{1}{\vert x_1-x'_1\vert^{\beta}} 
\theta(\sigma-\vert x_1-x'_1\vert) \nonumber \\
& \sim & I_2(0) - 2 L \int_0^{\sigma} {\rm d}x\, x^{-\beta}
= I_2(0) - 2 L \frac{\sigma^{1-\beta}}{1-\beta} . 
\end{eqnarray} 
For a general integral $I_{2N}(\sigma)$, we expand the product of theta 
functions in (\ref{confint}) as follows 
\begin{eqnarray}
\prod_{i,j=1}^N \theta(\vert x_i-x'_j\vert-\sigma) & = &
1 - \sum_{i,j=1}^N \theta(\sigma-\vert x_i-x'_j\vert) \nonumber \\ 
& & + \sum_{i,j,k,l=1\atop (i,j)\ne (k,l)}^N 
\theta(\sigma-\vert x_i-x'_j\vert) \theta(\sigma-\vert x_k-x'_l\vert) + \cdots .
\nonumber \\ & & \label{general}
\end{eqnarray}
The first term on the rhs (unity) gives $I_{2N}(0)$.
The second term contains $N^2$ summands, each provides the same contributions 
because the Boltzmann weight (\ref{Boltzmann}) is symmetric with respect to 
any interchange of two $x_i$ or two $x'_i$. 
This is why we can substitute the sum by
$N^2 \theta(\sigma-\vert x_1-x'_1\vert)$. 
The oppositely charged particles at $x_1$ and $x'_1$ have to be very close
to one another for small $\sigma$, and therefore form an almost neutral entity
which decouples from all other charges.
The integration over $x_1$ and $x'_1$  produces the previous factor 
$- 2 L \sigma^{1-\beta}/(1-\beta)$, while the integration over
$x$-coordinates of all remaining $2(N-1)$ charges, when multiplied by 
$N^2/(N!)^2$, implies $I_{2(N-1)}(0)$.
The summands in the third term on the rhs of (\ref{general}) can have
$i=k$ or $j=l$; we exclude such terms because they describe the situation
of one say $(+)$ charge coupled to two $(-)$ charges and the corresponding
power $\sigma^{2-\beta}$ is marginal in comparison with $\sigma^{1-\beta}$.
Then there is $[N(N-1)]^2/2!$ of equivalent terms of type 
$\theta(\sigma-\vert x_1-x'_1\vert) \theta(\sigma-\vert x_2-x'_2\vert)$.
As before, the couples of particles with coordinates $(x_1,x'_1)$ and
$(x_2,x'_2)$ form neutral entities for small $\sigma$, each contributing
by the factor $- 2 L \sigma^{1-\beta}/(1-\beta)$.
The integration over $x$-coordinates of all remaining $2(N-2)$ charges, 
multiplied by $[N(N-1)]^2/(2!N!^2)$, implies $I_{2(N-2)}(0)/2!$.
Proceeding further in this way we end up with the recurrence relation
\begin{equation}
I_{2N}(\sigma) = I_{2N}(0) - 
\left( 2 L \frac{\sigma^{1-\beta}}{1-\beta} \right) I_{2(N-1)}(0)
+ \frac{1}{2!} \left( 2 L \frac{\sigma^{1-\beta}}{1-\beta} \right)^2 
I_{2(N-2)}(0) + \cdots  
\end{equation} 
with $I_0(0)=1$.
Inserting this recurrence into the definition (\ref{xinew}), we end up with
\begin{equation}
\Xi_L(z,\sigma) = \Xi_L(z,0) 
\exp\left( - 2 z^2 \frac{\sigma^{1-\beta}}{1-\beta} L \right) .  
\end{equation}
Applying the logarithm to both sides of this equation and using 
the definition of the pressure (\ref{defpressure}), we recover 
the needed result (\ref{resultpressure}).

\renewcommand{\theequation}{3.\arabic{equation}}
\setcounter{equation}{0}

\section{Thermodynamics of the ordered 1D log-gas} \label{Sect.3}
We proceed by the log-gas system with alternating $\pm$ charges.
To distinguish the quantities from the unconstrained case, we add to
them the upperscript ``(ord)'', i.e. ordered.
As before, we start with pointlike charges in Sect. \ref{sub1} and then, 
in order to pass the collapse point $\beta_c=1$, we attach to particles 
a small hard core in Sect. \ref{sub2}. 

\subsection{Pointlike particles} \label{sub1}
For a neutral configuration of $2N$ particles constrained to a circle 
of circumference $L$, we consider only ordered sequences with alternating
$\pm$ charges: 
\begin{equation} \label{ordconf}
\varphi_1> \varphi'_1> \varphi_2> \varphi'_2 \cdots
> \varphi_N> \varphi'_N .
\end{equation}
The grand partition function is expressible as
\begin{equation} \label{Xiord}
\Xi^{(\rm ord)}_L(z) = 1 + \sum_{N=1}^{\infty} 
\left[ (2\pi)^{\beta/2} z L^{1-\frac{\beta}{2}} \right]^{2N} I^{\rm(ord)}_{2N} ,
\end{equation}
where the configuration integrals
\begin{eqnarray} 
I^{\rm(ord)}_{2N} & = & \int_0^{2\pi} \frac{{\rm d}\varphi_1}{2\pi} 
\int_0^{\varphi_1} \frac{{\rm d}\varphi'_1}{2\pi} 
\int_0^{\varphi'_1} \frac{{\rm d}\varphi_2}{2\pi} 
\int_0^{\varphi_2} \frac{{\rm d}\varphi'_2}{2\pi} \cdots
\int_0^{\varphi_{N-1}} \frac{{\rm d}\varphi_N}{2\pi} 
\int_0^{\varphi_N} \frac{{\rm d}\varphi'_N}{2\pi} \nonumber \\ 
& & \times B_{2N}(\{ \varphi_i\},\{ \varphi'_i\}) \label{Iord}  
\end{eqnarray}
involve the Boltzmann weights (\ref{Boltzmannweight}).
We are interested in the thermodynamic limit of the pressure $P^{\rm(ord)}$
and the total particle density $n^{\rm(ord)}$,
\begin{equation} 
\beta P^{\rm(ord)}(z) = \lim_{L\to\infty} \frac{1}{L} \ln \Xi_L^{\rm(ord)}(z) ,
\qquad n^{\rm(ord)}(z) = z \frac{\partial}{\partial z} \beta P^{\rm(ord)}(z) .
\end{equation}
As in the unconstrained case, both $\beta P^{\rm(ord)}(z)$ and $n^{\rm(ord)}(z)$ 
scale with the fugacity like $z^{\frac{2}{2-\beta}}$.

The ordered configuration integrals (\ref{Iord}) can be compared to
their unconstrained counterparts (\ref{defI}).
The point is that the Boltzmann weights (\ref{Boltzmannweight}) are invariant 
with respect to any interchange of two $\varphi_i$ or two $\varphi'_i$.
Since there exist $N!$ possible ways how to order $\{ \varphi_i\}$,
and similarly for $\{ \varphi'_i\}$, we can choose the special angle 
constraints 
\begin{equation} \label{conf}
\varphi_1> \varphi_2> \cdots > \varphi_{N-1} > \varphi_N , \qquad
\varphi'_1> \varphi'_2> \cdots > \varphi'_{N-1} > \varphi'_N
\end{equation}
and rewrite (\ref{defI}) as follows 
\begin{eqnarray}
I_{2N} & = & \int_0^{2\pi} \frac{{\rm d}\varphi_1}{2\pi} 
\int_0^{2\pi} \frac{{\rm d}\varphi'_1}{2\pi} 
\int_0^{\varphi_1} \frac{{\rm d}\varphi_2}{2\pi} 
\int_0^{\varphi'_1} \frac{{\rm d}\varphi'_2}{2\pi} \cdots
\int_0^{\varphi_{N-1}} \frac{{\rm d}\varphi_N}{2\pi} 
\int_0^{\varphi'_{N-1}} \frac{{\rm d}\varphi'_N}{2\pi} \nonumber \\ 
& & \times B_{2N}(\{ \varphi_i\},\{ \varphi'_i\}) . 
\end{eqnarray}
The alternating configuration space (\ref{ordconf}) constitutes a subset of
the reduced configuration space (\ref{conf}).
Since the Boltzmann weights are positive, we conclude that
$I_{2N} \ge I_{2N}^{\rm(ord)}$ or, equivalently, 
$\Xi_L(z) \ge \Xi_L^{\rm(ord)}(z)$. 
In the thermodynamic limit $L\to\infty$, we have the rigorous inequalities
\begin{equation} \label{ineq}
\beta P(z) \ge \beta P^{\rm(ord)}(z) , \qquad n(z) \ge n^{\rm(ord)}(z) .
\end{equation} 

We take the TBA results from \cite{Fendley95a}. 
The expansion parameter in (\ref{Xiord}) is expressible in terms of 
the length scale $L_{\rm B}$ in analogy with (\ref{exp}) if the substitution
$z\to \tau z$ with $\tau={\rm i}/(q-q^{-1}) = 1/[2\sin(\pi\beta/2)]$ is made:
\begin{equation}
\frac{z}{2\sin\left(\frac{\pi\beta}{2}\right)} 
(2\pi)^{\beta/2} L^{1-\frac{\beta}{2}} = \Gamma\left(\frac{\beta}{2}\right)
\left[ \frac{L}{L_{\rm B}} \frac{\Gamma\left(\frac{1}{2-\beta}\right)}{
2\sqrt{\pi}\Gamma\left(\frac{\beta}{2(2-\beta)}\right)} 
\right]^{1-\frac{\beta}{2}} .
\end{equation}
For large $L$, the grand partition function for the ordered system
was found to behave as
\begin{equation}
\Xi^{\rm(ord)}_L \mathop{\sim}_{L\to\infty} \exp\left[ \frac{L}{L_{\rm B}} 
\tan\left(\frac{\pi\beta}{2(2-\beta)}\right) \right] .
\end{equation}
Eliminating $L_{\rm B}$ from the two equations and using the formulas 
(\ref{Gamma}) for the Gamma functions, the bulk pressure is obtained in 
the form
\begin{equation} \label{pressord}
\beta P^{\rm(ord)}(z) = \frac{1}{\sqrt{\pi}} 
\frac{\Gamma\left( \frac{1-\beta}{2-\beta} \right)}{
\Gamma\left( \frac{4-3\beta}{4-2\beta} \right)} 
\left[ \Gamma\left( 1-\frac{\beta}{2} \right) z \right]^{\frac{2}{2-\beta}} .
\end{equation}

\begin{figure}[h] \label{fig3} 
\begin{center}
\includegraphics[clip,width=0.85\textwidth]{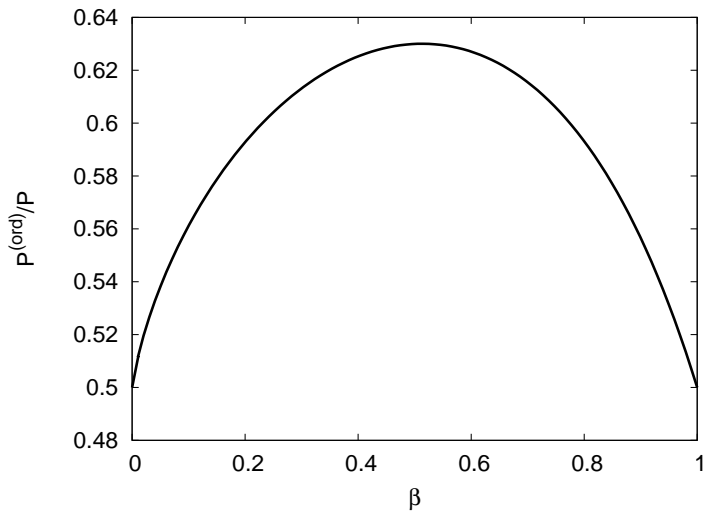} 
\end{center}
\caption{The ratio of the pressures for the ordered and unconstrained
charge systems versus the inverse temperature $\beta$.}
\end{figure}

For a fixed $z$, the ratio of this pressure to the pressure for the 
unconstrained charge system (\ref{betaP}) turns out to be
\begin{equation} \label{pressureratio}
\frac{P^{\rm(ord)}}{P} = 2 \sin\left( \frac{\pi\beta}{2(2-\beta)}\right)
\left[ \frac{1}{2\sin\left(\frac{\pi\beta}{2}\right)} 
\right]^{\frac{2}{2-\beta}} .
\end{equation}
This ratio is plotted as a function of the inverse temperature $\beta$
in Fig. 3.
It is seen that the pressure inequality in (\ref{ineq}) is satisfied.
Interestingly, the ratio takes the minimum value $1/2$ just at
the limiting points $\beta\to 0$ and $\beta\to 1$.

In Appendix A, a fusion relation between the grand partition functions of 
the unconstrained and ordered log-gases is used to rederive in an alternative 
way the relationship between the corresponding bulk pressures 
(\ref{pressureratio}).

Having the bulk pressure (\ref{pressord}), the density-fugacity relationship 
reads as
\begin{equation} \label{nz2}
\frac{n^{\rm(ord)}}{z^{\frac{2}{2-\beta}}} = \frac{2}{\sqrt{\pi}(2-\beta)} 
\frac{\Gamma\left( \frac{1-\beta}{2-\beta} \right)}{
\Gamma\left( \frac{4-3\beta}{4-2\beta} \right)} 
\left[ \Gamma\left( 1-\frac{\beta}{2} \right) \right]^{\frac{2}{2-\beta}} .
\end{equation}
The small-$\beta$ expansion of the rhs of this formula 
\begin{equation} \label{smallbord}
\frac{n^{\rm(ord)}}{z^{\frac{2}{2-\beta}}} = 1 + \frac{1}{2} (C+1+\ln 2) \beta
+ O(\beta^2)
\end{equation}
is analytic, which is in contrast to the singular result (\ref{smallb}) 
for the unconstrained log-gas.
This expansion is checked in Appendix B, using pair distributions of 
the free $(\beta=0)$ log-gas with alternating $\pm$ charges \cite{Alastuey95}.
For a fixed $z$ and in the limit $\beta\to 1^-$, the term 
$\Gamma((1-\beta)/(2-\beta))\sim 1/(1-\beta)$ implies the collapse singularity
\begin{equation} \label{singord}
n^{\rm(ord)} \mathop{\sim}_{\beta\to 1^-} \frac{2 z^2}{1-\beta} . 
\end{equation} 
The adequacy of this term, in the sense that it compensates exactly 
the singularity of the hard-core contribution at the collapse point, 
will be documented in the next part.  
The plot of the density-fugacity relationship for ordered pointlike particles
(\ref{nz2}), valid up to $\beta_c=1$, is represented in Fig. 4 
by the solid curve. 
The curve is monotonously increasing in the whole $\beta$-interval $[0,1]$,
which is in contrast to Fig. 2 for the unconstrained system. 

\begin{figure}[h] \label{fig4} 
\begin{center}
\includegraphics[clip,width=0.85\textwidth]{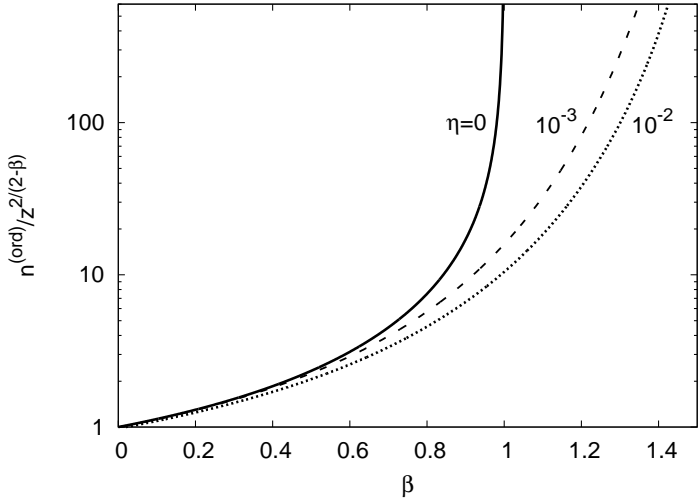} 
\end{center}
\caption{The density-fugacity relationship versus the inverse temperature 
$\beta$ for the ordered log-gas.
The solid curve describes pointlike charges with the packing fraction 
$\eta=0$. 
The dashed and dotted curves correspond to $\eta=10^{-3}$ and $10^{-2}$, 
respectively.}
\end{figure}

The virial equation of state takes the usual form
\begin{equation}
\beta P^{\rm(ord)} = \left( 1 - \frac{\beta}{2} \right) n^{\rm(ord)} . 
\end{equation}
The (dimensionless) Helmholtz free energy per particle reads
\begin{eqnarray}
f^{\rm(ord)}(\beta,n) & = & \ln(\lambda n) - \frac{\beta}{2} \ln(r_0 n)
+ \frac{1}{2} \left( 1 - \frac{\beta}{2}\right) \ln\pi \nonumber \\ & & 
- \left( 1 - \frac{\beta}{2} \right)  
\left[ 1 - \ln\left( 1 - \frac{\beta}{2} \right) \right]
- \ln\Gamma\left( 1-\frac{\beta}{2}\right) 
\nonumber \\ & & 
- \left( 1 - \frac{\beta}{2} \right)  \left[
\ln\Gamma\left( \frac{1-\beta}{2-\beta} \right)
- \ln\Gamma\left( \frac{4-3\beta}{4-2\beta} \right) \right] .
\end{eqnarray}
For the excess specific heat, we obtain
\begin{eqnarray}
\frac{c_v^{\rm(ord)ex}}{k_{\rm B}} & = & \frac{\beta^2}{2(2-\beta)^3}
\left[ \psi'\left(\frac{1-\beta}{2-\beta}\right) -
\psi'\left(\frac{4-3\beta}{4-2\beta)}\right) \right] \nonumber \\
& & - \frac{\beta^2}{2(2-\beta)} 
+ \frac{\beta^2}{4} \psi'\left(1-\frac{\beta}{2}\right)  .
\end{eqnarray}
Close to the collapse point $\beta\to 1^-$, it exhibits the singular
behavior 
\begin{equation}
\frac{c_v^{\rm(ord)ex}}{k_{\rm B}} = \frac{1}{2(1-\beta)^2}
- \frac{3}{2(1-\beta)} + O(1) ,
\end{equation}
which is exactly the same as in the unconstrained case, 
see Eq. (\ref{specsing}).

\subsection{Particles with hard cores} \label{sub2}
We attach to particles hard cores of diameter $\sigma$ and derive 
the equation of state directly from the definition of the grand partition 
function, in close analogy with the procedure for the unconstrained system 
starting from Eq. (\ref{resultpressure}).

For the ordered charge system of large size $L$, the grand partition function
is defined by
\begin{equation} \label{xineword} 
\Xi_L^{\rm(ord)}(z,\sigma) = 
1 + \sum_{N=1}^{\infty} z^{2N} I_{2N}^{\rm(ord)}(\sigma) ,
\end{equation}
where the configuration integrals
\begin{eqnarray} 
I_{2N}^{\rm(ord)}(\sigma) & = & \int_0^L {\rm d}x_1 \int_0^{x_1} {\rm d}x'_1
\int_0^{x'_1} {\rm d}x_2 \int_0^{x_2} {\rm d}x'_2 \cdots 
\int_0^{x'_{N-1}} {\rm d}x_N \int_0^{x_N} {\rm d}x'_N \nonumber \\ & & \times
B_{2N}(\{ x_i\},\{ x'_i\}) \prod_{i=1}^N \theta(x_i-x'_i-\sigma)
\prod_{i=1}^{N-1} \theta(x'_i-x_{i+1}-\sigma) \phantom{aaaa} \label{confintord}
\end{eqnarray}
involve the Boltzmann weights (\ref{Boltzmann}).
As before, we substitute each Heaviside theta function in (\ref{confintord}) 
via (\ref{trick}).
For the lowest-order integral, we get
\begin{eqnarray}
I_2^{\rm(ord)}(\sigma) & = & I_2^{\rm(ord)}(0) - \int_0^L {\rm d}x_1 \int_0^{x_1} 
{\rm d}x'_1 \frac{1}{\vert x_1-x'_1\vert^{\beta}} 
\theta(\sigma-x_1+x'_1) \nonumber \\
& \sim & I_2(0) - L \int_0^{\sigma} {\rm d}x\, x^{-\beta}
= I_2(0) - L \frac{\sigma^{1-\beta}}{1-\beta} . 
\end{eqnarray} 
For a general integral $I_{2N}^{\rm(ord)}$ in (\ref{confintord}), 
we expand the product of theta functions written as in (\ref{trick}).
The first term (unity) implies $I_{2N}^{\rm(ord)}(0)$.
The second term contains $(2N-1)$ theta functions; each theta function forces 
two oppositely charged nearest-neighbor particles to be very close to one 
another for small $\sigma$, and therefore to form an almost neutral 
entity which decouples from all other charges.
The integration over distance between these nearest neighbors produces 
the factor $-\sigma^{1-\beta}/(1-\beta)$. 
The integration over the $x$-coordinate of the neutral entity, 
which exhibits all possible orderings with respect to a given configuration 
of other ``active'' charges, implies $L$.
The remaining $2(N-1)$ charges contribute by $I_{2(N-1)}^{\rm(ord)}(0)$, 
so the total contribution is
$[- L \sigma^{1-\beta}/(1-\beta)]I_{2(N-1)}^{\rm(ord)}(0)$.
In the case of $k$ neutral entities, there is an indistinguishability
factor $1/k!$ attached to $[- L \sigma^{1-\beta}/(1-\beta)]^k
I_{2(N-k)}^{\rm(ord)}(0)$.
We end up with the recurrence relation
\begin{equation}
I_{2N}^{\rm(ord)}(\sigma) = I_{2N}^{\rm(ord)}(0) - 
\left( L \frac{\sigma^{1-\beta}}{1-\beta} \right) I_{2(N-1)}^{\rm(ord)}(0)
+ \frac{1}{2!} \left( L \frac{\sigma^{1-\beta}}{1-\beta} \right)^2 
I_{2(N-2)}^{\rm(ord)}(0) + \cdots  
\end{equation} 
with $I_0^{\rm(ord)}(0)=1$.
Inserting this recurrence into the definition (\ref{xineword}), we end up with
\begin{equation}
\Xi_L^{\rm(ord)}(z,\sigma) = \Xi_L^{\rm(ord)}(z,0) 
\exp\left( - z^2 \frac{\sigma^{1-\beta}}{1-\beta} L \right) .  
\end{equation}
Applying the logarithm to both sides of this equation and in the limit
$L\to\infty$, we get the bulk pressure of the ordered system of charges
with hard cores:
\begin{equation} \label{resultpressureord}
\beta P^{\rm(ord)}(z,\sigma) = \beta P^{\rm(ord)}(z,0) 
- z^2 \frac{\sigma^{1-\beta}}{1-\beta} , \qquad \xi\to 0 ,
\end{equation}
where $\beta P^{\rm(ord)}(z,0)$ is given by (\ref{pressord}).
The corresponding density-fugacity relationship reads as
\begin{equation} \label{basicord}
n^{\rm(ord)}(z,\sigma) = n^{\rm(ord)}(z,0) 
- 2 z^2 \frac{\sigma^{1-\beta}}{1-\beta} , \qquad \xi \to 0 .
\end{equation}
Although these results were derived in the stability region $0\le \beta <1$,
we assume that they can be analytically continued beyond the collapse 
point $\beta_c=1$.

To provide a convenient representation of (\ref{basicord}), we express 
the density-fugacity relation for pointlike particles (\ref{nz2}) as follows
\begin{equation} \label{pointdensityord}
n^{\rm(ord)}(z,0) = \frac{2\Phi^{\rm(ord)}(\beta)}{1-\beta} z^{\frac{2}{2-\beta}} ,
\end{equation}
where the function
\begin{equation} \label{Phiord}
\Phi^{\rm(ord)}(\beta) = \frac{1}{\sqrt{\pi}} 
\frac{\Gamma\left( \frac{3-2\beta}{2-\beta} \right)}{
\Gamma\left( \frac{4-3\beta}{4-2\beta} \right)}
\left[ \Gamma\left( 1-\frac{\beta}{2} \right) \right]^{\frac{2}{2-\beta}} 
\end{equation}
was chosen such that it equals to 1 at the collapse point, namely
\begin{equation}
\Phi^{\rm(ord)}(\beta) = 1 + \left( C +\ln\pi \right) (\beta-1) 
+ O((\beta-1)^2) .
\end{equation}
Using the representation (\ref{pointdensityord}) in (\ref{basicord}), we get
the density-fugacity relationship which involves the leading hard-core 
correction:
\begin{equation} \label{hcresultord}
\frac{n^{\rm(ord)}(z,\sigma)}{z^{\frac{2}{2-\beta}}} = \frac{2}{1-\beta}
\left[ \Phi^{\rm(ord)}(\beta) - \xi^{1-\beta} \right] , \qquad \xi\to 0 .
\end{equation}
The leading correction $\propto \xi^{1-\beta}$ makes the density finite at 
the collapse point $\beta_c=1$,
\begin{equation}
\frac{n^{\rm(ord)}(z,\sigma)}{z^2} = - 2 \left( \ln\xi + C + \ln\pi \right) .
\end{equation}
The formalism holds up to $\beta=3/2$ at which $\Phi^{\rm(ord)}(\beta)$ diverges.

It is convenient to pass from the parameter $\xi=\sigma z^{\frac{2}{2-\beta}}$ 
to the packing fraction $\eta = n^{\rm(ord)} \sigma$.
Denoting $g\equiv n^{\rm(ord)}(z,\sigma)/z^{\frac{2}{2-\beta}}$, we have $\eta= g\xi$.
Multiplying both sides of Eq. (\ref{hcresultord}) by 
$(1-\beta) g^{1-\beta}$, we get
\begin{equation}
(1-\beta) g^{2-\beta} - 2 \left[ \Phi^{\rm(ord)}(\beta) g^{1-\beta} 
- \eta^{1-\beta} \right] = 0 .
\end{equation}
At $\beta=0$, we take the solution $g=(1+\sqrt{1-8\eta})/2$ 
and follow this branch with increasing $\beta$.
The density-fugacity relation for the packing fractions $10^{-3}$ and
$10^{-2}$ are plotted in Fig. 4 by the dashed and dotted curves, 
respectively.
The functions are monotonously increasing in the whole interval of $\beta$,
which is in contrast to the unconstrained system (see Fig. 2).

\renewcommand{\theequation}{4.\arabic{equation}}
\setcounter{equation}{0}

\section{Conclusion} \label{Sect.4}
The aim of this work was to derive and analyze the exact thermodynamics
of the two-component log-gas formulated on an infinite line with 
alternating $\pm$ charges.

The thermodynamics was first solved for the case of pointlike particles,
up to the collapse point $\beta_c=1$, by using the TBA results. 
The obtained density-fugacity relationship (\ref{nz2}) turns out to be
analytic in the high-temperature region, see the small-$\beta$ expansion 
(\ref{smallbord}) and its check in the Appendix B.
Moreover, as is shown in Fig. 4 by the solid curve, the density-fugacity
ratio $n^{\rm(ord)}/z^{2/(2-\beta)}$ is a monotonously increasing function
of the inverse temperature $\beta$. 
This is in contrast to the unconstrained log-gas with a non-analytic
small-$\beta$ expansion and a non-monotonous plot of $n/z^{2/(2-\beta)}$
versus $\beta$, see the solid curve in Fig. 2.
For a fixed fugacity $z$, the validity of the inequality (\ref{ineq}) 
for the pressures of the unconstrained and ordered log-gases was
confirmed by finding the exact ratio $P^{\rm(ord)}/P$ (Fig. 3); 
the minimum $1/2$ is reached just at the limiting points $\beta\to 0$ 
and $\beta\to 1$.

The inclusion of hard cores around the particles was a more complicated
task than in the unconstrained case since we could not apply 
the screening sum rule for pair correlation functions valid for a conductor.
The derivation of the leading hard-core correction to the pressure, 
outlined in Sect. 3.2, was based on the explicit definition of 
the grand partition function and a substitution of Heaviside theta 
functions due to the presence of hard cores.
The final result for the bulk pressure (\ref{resultpressureord}) and 
the corresponding density-fugacity relationship (\ref{basicord}) 
pass the collapse test, in the sense that the hard-core correction 
compensates exactly the singularity of the pointlike model and leads 
to a finite pressure and particle density at $\beta=1$. 
 
Since the log-gas is by the structure much simpler that the 2D Coulomb gas, 
it might be possible to get all relevant hard-core corrections which 
compensate an infinite series of singularities of the pointlike result 
and allow us to go up to $\beta_{\rm KT}=2$, in the spirit of Refs. 
\cite{Kalinay02,Tellez05,Tellez07}.
This problem is left for future.

\renewcommand{\theequation}{A.\arabic{equation}}
\setcounter{equation}{0}

\section*{Appendix A} \label{AppendixA}
Here, we rederive in an alternative way the relationship between the bulk 
pressures of the ordered and unconstrained log-gases (\ref{pressureratio}). 

We use the fusion relation between the grand partition functions of 
the unconstrained and ordered log-gases \cite{Fendley95c,Fendley96}:
\begin{equation} \label{fusion}
\Xi_L^{\rm(ord)}((q-q^{-1})z) = \frac{\Xi_L(qz)+\Xi_L(q^{-1}z)}{2\Xi_L(z)} ,
\qquad q={\rm e}^{{\rm i}\pi\beta/2} .
\end{equation}
We know that for large $L$ the grand partition functions behave as
\begin{equation}
\Xi_L^{\rm(ord)}(z) \mathop{\sim}_{L\to\infty} 
\exp\left[ \beta P^{\rm(ord)}(z) L \right] , \qquad 
\beta P^{\rm(ord)}(z) = a^{\rm(ord)}(\beta) z^{\frac{2}{2-\beta}} 
\end{equation}
and 
\begin{equation}
\Xi_L(z) \mathop{\sim}_{L\to\infty} \exp\left[ \beta P(z) L \right] ,
\qquad \beta P(z) = a(\beta) z^{\frac{2}{2-\beta}} .
\end{equation}
The lhs of Eq. (\ref{fusion}) reads
\begin{eqnarray} 
& &\exp\left\{ - L a^{\rm(ord)}(\beta) 
\left[ 2 z \sin\left( \frac{\pi\beta}{2} \right) \right]^{\frac{2}{2-\beta}}
\sin\left(\frac{\pi\beta}{2(2-\beta)}\right) \right\}  
\nonumber \\ & & \quad \times
\cos\left\{ L a^{\rm(ord)}(\beta) 
\left[ 2 z \sin\left( \frac{\pi\beta}{2} \right) \right]^{\frac{2}{2-\beta}}
\cos\left(\frac{\pi\beta}{2(2-\beta)}\right) \right\} . \label{eq1}
\end{eqnarray}
The rhs of Eq. (\ref{fusion}) can be expressed as 
\begin{equation} \label{eq2}
\exp \left\{ L a(\beta) z^{\frac{2}{2-\beta}} \left[
\cos\left(\frac{\pi\beta}{2-\beta}\right) -1 \right] \right\}
\cos\left[ L a(\beta) z^{\frac{2}{2-\beta}} 
\sin\left(\frac{\pi\beta}{2-\beta}\right) \right] . 
\end{equation} 
The equality of separately exponential and cosine terms in Eqs. (\ref{eq1}) 
and (\ref{eq2}) leads to the only relation
\begin{equation} \label{pressurerationew}
\frac{a^{\rm(ord)}(\beta)}{a(\beta)} = \frac{P^{\rm(ord)}}{P} 
= 2 \sin\left( \frac{\pi\beta}{2(2-\beta)}\right)
\left[ \frac{1}{2\sin\left(\frac{\pi\beta}{2}\right)} 
\right]^{\frac{2}{2-\beta}} ,
\end{equation}
in agreement with (\ref{pressureratio}).

\renewcommand{\theequation}{B.\arabic{equation}}
\setcounter{equation}{0}

\section*{Appendix B} \label{AppendixB}
Here, we construct the small-$\beta$ expansion of the pressure and
the particle density for the ordered log-gas.

For $\beta=0$, the Boltzmann factor 
$B_{2N}(\{ \varphi_i\},\{\varphi'_i\}) = 1$.
The ordered integrals (\ref{Iord}) are simply given by
$I_{2N}^{{\rm(ord)}}=1/(2N)!$ as there are just $(2N)!$ ways how to order 
the angle variables and each ordering implies the same contribution.
This leads to
\begin{equation}
\Xi_L^{{\rm(ord)}} = 1 + \sum_{N=1}^{\infty} \frac{(zL)^{2N}}{(2N)!}
\mathop{\sim}_{L\to\infty} \frac{1}{2} \exp(z L) .
\end{equation} 
Consequently, in the lowest order,
\begin{equation} \label{lowest}
\beta P_0^{\rm(ord)}(z) = z , \qquad n_0^{\rm(ord)}(z) = z .
\end{equation}

For a given configuration of particles $\{ j\}$ with charges $\{ q_j\}$
at positions $\{ x_j\}$, the interaction energy is given by
\begin{eqnarray}
E & = & \frac{1}{2} \sum_{j\ne k} q_j q_k (-\ln\vert x_j-x_k\vert) \nonumber \\
& = & \frac{1}{2} \int_0^L {\rm d}x \int_0^L {\rm d}x'\, (-\ln\vert x-x'\vert) 
\sum_{q,q'=\pm 1} q q' \hat{n}_{qq'}(x,x') , 
\end{eqnarray}
where the microscopic quantity
\begin{equation}
\hat{n}_{qq'}(x,x') = \sum_{j\ne k} \delta_{q,q_j} \delta(x-x_j)
\delta_{q',q_k} \delta(x'-x_k) .
\end{equation}
Expanding the Boltzmann factor in $\beta$, 
${\rm e}^{-\beta E} \sim 1 - \beta E$, and using the cumulant expansion,
we get the leading $\beta$-correction to the pressure,
\begin{equation}
\beta P_1^{\rm(ord)}(z) = \beta P_0^{\rm(ord)}(z) 
- \beta \frac{\langle E \rangle_0}{L} , 
\end{equation}
where the thermal averaging $\langle \cdots \rangle_0$ is over the system of
non-interacting $(\beta=0)$ ordered charges.
For an infinite system $L\to\infty$, the two-body distributions $n_{qq'}(x,x') 
\equiv n_{qq'}(\vert x-x'\vert) = \langle \hat{n}_{qq'}(x,x') \rangle_0$ 
have been calculated in \cite{Alastuey95}:
\begin{equation}
n_{qq}(x) = \left( \frac{n}{2}\right)^2 
\left( 1 - {\rm e}^{-2n\vert x\vert} \right) , \qquad
n_{q,-q}(x) = \left( \frac{n}{2}\right)^2 
\left( 1 + {\rm e}^{-2n\vert x\vert} \right) .
\end{equation}
Considering the lowest order for $n$ from (\ref{lowest}), we obtain
\begin{eqnarray}
\beta P_1^{\rm(ord)}(z) & = & z 
+ \frac{\beta}{2} \int_{-\infty}^{\infty} {\rm d}x\, \ln\vert x\vert 
\left[ n_{++}(x) + n_{--}(x) - n_{+-}(x) - n_{-+}(x) \right] \nonumber \\
& = & z + \frac{\beta}{2} z ( C + \ln 2) + \frac{\beta}{2} z \ln z .
\end{eqnarray}
The corresponding density of particles
\begin{eqnarray}
n_1^{\rm(ord)}(z) = z + \frac{\beta}{2} z ( C + 1 + \ln 2) 
+ \frac{\beta}{2} z \ln z
\end{eqnarray}
is consistent with the small-$\beta$ expansion (\ref{smallbord}),
where $z^{\frac{2}{2-\beta}} \sim z[1+(\beta/2)\ln z]$.

To find the next order of the small-$\beta$ expansion we need to know
four-body distributions for the system of non-interacting ordered charges,
and so on.

\begin{acknowledgements}
I am grateful to Peter Forrester for directing my attention to works 
about equivalence between the grand partition functions of unconstrained 
and ordered log-gases. 
The support received from Grant VEGA No. 2/0049/12 is acknowledged. 
\end{acknowledgements}

\end{document}